# Electron Driven Mobility Model by Light on the Stacked Metal-Dielectric-Interfaces

N. Pornsuwancharoen[1], P. Youplao[1], I.S. Amiri[2], J. Ali [3,4] and P. Yupapin[5,6*]

[1]Department of Electrical Engineering, Faculty of Industry and Technology,
Rajamangala University of Technology Isan, Sakon Nakhorn Campus, Sakon Nakhorn, Thailand;
E-mails: jeewuttinun@gmail.com; pichai3112@yahoo.com
[2]Photonics Research Centre, University of Malaya, 50603 Kuala Lumpur, Malaysia;
E-mail: irajsadeghamiri@gmail.com
[3]Laser Centre, IBNU SINA ISIR, Universiti Teknologi Malaysia 81310 Johor Bahru, Malaysia;
[4]Faculty of Science, Universiti Teknologi Malaysia, 81310 Johor Bahru, Malaysia; E-mail:djxxx_1@yahoo.com
[5]Department for Management of Science and Technology Development, Ton Duc Thang University,
District 7, Ho Chi Minh City, District 7, Vietnam;
[6]Faculty of Electrical & Electronics Engineering, Ton Duc Thang University,
District 7, Ho Chi Minh City, Vietnam;
*Corresponding author E-mail: <preecha.yupapin@tdt.edu.vn>

**Abstract:** An electron mobility enhancement is the very important phenomenon of an electron in the electronic device, where the high electronic device performance has the good electron mobility, which is obtained by the overall electron drift velocity in the electronic material driven potential difference. The increasing in electron mobility by the injected high group velocity pulse is proposed in this article. By using light pulse input into the nonlinear microring resonator, light pulse group velocity can be tuned and increased, from which the required output group velocity can be obtained, which can be used to drive electron within the plasmonic waveguide, where eventually, the relative electron mobility can be obtained, the increasing in the electron mobility after adding up by the driven optical fields can be connected to the external electronic devices and circuits, which can be useful for many applications.

## 1. Introduction

An electron mobility is the basic property of all electronic devices and circuits, in which the drift velocity of an electron is brought to have the device or circuit performance, where the increasing in an electron speed within the conductor(semiconductor) is required, which can be realized by using light-on-metal-dielectric-interfaces, from which the speed of electron can be driven relatively increase before entering into the electronic circuits, where the shorter reaching time between anode and cathode is the major parameter, leads the output current is increased. The mobility of electron in silicon and GaAs are 1900 and 4500-6500 $cm^{-2}\ V^{-1}s^{-1}$, respectively [1-4], while the electron mobility in suspended graphene is $2 \times 10^5\ cm^{-2}\ V^{-1}s^{-1}$. In this proposal, the group velocity of the driven fields is added up to the electron drift velocity on the stacked gold atom-silicon-graphene, where the greater relative electron mobility can be obtained. Moreover, the circuit noise reduction can also be obtained by using a single electron device, for example, a gold atom, from which the electron mobility noise effect from the electron collision may be neglected, which may be used such a concept for room temperature superconductor. In application, the stacked waveguide can be constructed (fabricated) to be a device or circuit, in which metals such as gold, silver, ferromagnetic, copper etc. can be formed to be a device or circuit, which can be used to support the hybrid electronics such as classical, plasmonic, spintronic and quantum electronics can be combined together, from which the electron charge and spin can be used as the conducting agents [5-10].

Ultrafast light has been successfully implemented and useful in many applications [11-15], where none of them has been used in a nano-electronics regime. The use of fast light for fast switching time was reported, from which the system was the nonlinear ring resonator, which is called a panda ring resonator. A Panda ring resonator proposed by Yupapin is a high-performance semiconductor laser device based on the nonlinear optics. By using a panda ring resonator structure, the speed of the group velocity could be increased and controlled by the controlling of the coupling device parameters. Although a Panda ring resonator has the various attractive possibilities for human society and healthcare devices, an experimental demonstration of the panda ring resonator has not been reported. This is because the integration technology of different-type semiconductors in submicron space is extremely difficult. In this article, the benefits of group velocity control are used to drive the electron in the plasmonic waveguide, in which the increasing in switching plasmonic pulse is introduced the driven electron speed in the gold atom line circuit. The electron is induced by the light-graphene interaction and driven by the plasmonic wave group velocity, where finally, the electron mobility is increased and calculated.

In application, the use of the stacked waveguide in the form of metal-dielectric interfaces waveguide is proposed for hybrid electronic applications, where the use of classical, plasmonic, spintronic and quantum electronics can be made and realized. The use of electron mobility enhancement is modeled by using the micro-ring system described, from which the optical field group velocities can be generated and controlled for electron mobility enhancement for hybrid electronic use. From which the single electron charge device and the circuit will be realistic, while the use of electron spin will lead to having the circuit without current flow but the spin wave instead makes the circuit operates, where there is no short circuit from current in this case. The process of spin up and down is also made the quantum computer being interesting and more realized. The use of electron spin has a wide range of application because the small scale led more applications possible. The enhanced electron mobility can be used as the spin injection source for spintronic electronic use, where the use of spin wave in ferromagnetic nanostructures can provide the nonlinear operations similar to the semiconductor panda ring resonator because of the analogy between the photon and magnon, which is the spin wave quantization, from which the high performance nonlinear optical and spintronic devices can be obtained. Namely, instead of an optical wave packet, we realize the panda-ring resonator by using spin-wave packet [9, 10].

## 2. Background

In Figure 1, the driven fields group velocity in term of wavelength (frequency) changes is related to the group velocity, which will be added up the electron drift velocity, the relative velocity is $v + v_{driven}$, therefore, the relative electron mobility in the electric potential E is given by $v + v_{driven}/displacement$, which are described two parts as followings, , where firstly, the group velocity of the driven fields, where the Gaussian input pulse from a semiconductor laser that can be the embedded device, is input into the nonlinear micro-ring resonator, from which the single broadening pulse is disturbed and chopped by the two nonlinear side rings, in which the number of output pulses can be desired and the narrower pulse widths obtained. Thus, the pulse switching time is shorter than the initial single input peak, which means that each peak can travel faster than the initial input pulse. In the calculation, the use of the average traveling velocity, called a pulse group velocity can be obtained based on the resonant conditions. The energy in terms of electrical fields will be switched and transferred to the stacked gold atom-silicon-graphene interfaces waveguide. Secondly, the driven electrons in graphene by the evanescent wave from silicon to graphene will be driven by the external driven potential difference between cathode and anode. By using a single gold atom conductor, a single electron device can also be employed. Generally, the driven group velocity of optical fields within the micro ring system can be derived and tuned by the external nonlinear coupling effects, where the external disturbance can be generated by the two nonlinear side rings, in which the construction and deduction interferences of the center group velocity signals can be obtained. The input broadening pulse can be chopped to be small pulse width signals, which is the ultra-fast switching in time, which means that the gap between the peaks around the center wavelength is very short in time. Thus, the speed of traveling pulse is increased ($v = \lambda \Delta f$), which can be used to drive the electron mobility for various application, especially, in the surface plasmon media.

When the optical field ($E_i$) is input into the system as shown in Figure 1, the output fields of all ports are given by the following forms in Table 1 [13].

$$\left|\frac{E_{drop}}{E_{in}}\right|^2 = \frac{\kappa_1 \kappa_3 (y_1)(y_2) e^{-\frac{\alpha}{2}L_1}}{(y_3)(y_1) - 2(x_3)(x_1)\sqrt{1-\kappa_1}(x_2)(x_4)\sqrt{1-\kappa_3} e^{-\frac{\alpha}{2}L_1} \cos(knL_1) + (1-\kappa_1)(y_2)(y_4)(1-\kappa_3) e^{-\alpha L_1}}$$

$$\left|\frac{E_{th}}{E_{in}}\right|^2 = \frac{(1-\kappa_1)(y_3)(y_1) - 2(x_1)(x_2)(x_3)(x_4)\sqrt{1-\kappa_1}\sqrt{1-\kappa_3} e^{-\frac{\alpha}{2}L_1} \cos(knL_1) + (y_2)(y_4)(1-\kappa_3) e^{-\alpha L_1}}{(y_3)(y_1) - 2(x_3)(x_1)\sqrt{1-\kappa_1}(x_2)(x_4)\sqrt{1-\kappa_3} e^{-\frac{\alpha}{2}L_1} \cos(knL_1) + (1-\kappa_1)(y_2)(y_4)(1-\kappa_3) e^{-\alpha L_1}}$$

Where the other forms of the used parameters are given in Table 1, $E_{in}$: input port optical field, $E_{Through}$: through port optical field, $E_{Drop}$: drop port optical field, $E_{Add}$; add port optical field, $E_{ii}$: circulation optical field, $\kappa_i$: coupling constants.

Table 1: Ring resonator fields and parameters

| The optical fields: $E_i$ | The specified constant quantities: $C_n, x_n, y_n$ |
|---|---|
| $E_1 = E_{in} j\sqrt{\kappa_1} + E_6 \sqrt{1-\kappa_1} e^{-\frac{\alpha L_1}{24} - jkn\frac{L_1}{4}}$ | $C_1 = 1 - \sqrt{1-\kappa_4} e^{-\frac{\alpha}{2}L_3 - jkn L_3}$ |
| $E_2 = E_3 j\sqrt{\kappa_2} e^{-\frac{\alpha}{2}L_2 - jknL_2} + E_1 \sqrt{1-\kappa_2} e^{-\frac{\alpha L_1}{24} - jkn\frac{L_1}{4}}$ | $C_2 = \sqrt{1-\kappa_2} - e^{-\frac{\alpha}{2}L_2 - jknL_2}$ |
| $E_3 = E_1 j\sqrt{\kappa_2} e^{-\frac{\alpha L_1}{24} - jkn\frac{L_1}{4}} + E_3 \sqrt{1-\kappa_2} e^{-\frac{\alpha}{2}L_2 - jknL_2}$ | $C_3 = 1 - \sqrt{1-\kappa_2} e^{-\frac{\alpha}{2}L_2 - jknL_2}$ |
| $E_4 = E_2 \sqrt{1-\kappa_3} e^{-\frac{\alpha L_1}{24} - jkn\frac{L_1}{4}}$ as $E_{add} = 0$ | $C_4 = \sqrt{1-\kappa_4} - e^{-\frac{\alpha}{2}L_3 - jknL_3}$ |
| $E_5 = E_4 j\sqrt{\kappa_4} e^{-\frac{\alpha L_1}{24} - jkn\frac{L_1}{4}} + E_5 \sqrt{1-\kappa_4} e^{-\frac{\alpha}{2}L_3 - jkn L_3}$ | $x_n = |C_n|$ |
| $E_6 = E_5 j\sqrt{\kappa_4} e^{-\frac{\alpha}{2}L_3 - jknL_3} + E_4 \sqrt{1-\kappa_4} e^{-\frac{\alpha L_1}{24} - jkn\frac{L_1}{4}}$ | $y_n = |C_n|^2$ |

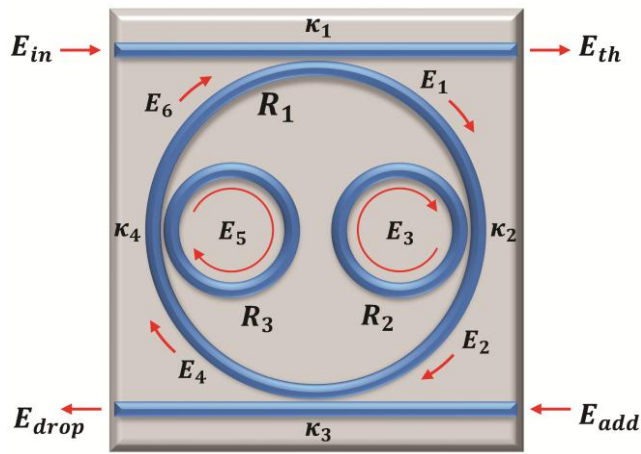

**Figure** 1: A micro-ring structure using a *GaAsInP/P* material, where the nonlinear coupling effect is introduced by the two nonlinear side rings

## 3. Results

From Figures 1 and 2, coherent light from the embedded source with the graphene active wavelength range is input into the system via the input port, from which the propagation of light is coupled into the center ring and two side rings, which is made by GaAsInP/P material. The selected parameters are given for demonstration. The required stable output optical fields at the resonant condition can be obtained, from which the required group velocity values for optically driven fields obtained. In practice, there are many pulse trains with different group velocities occurred, which can be selected by switching operation to connect with the stacked waveguide is required for the group velocity selected fields. The required group velocity can be measured by using the reference signal at the Drop port before the output at the through port being switched and used for the required driven electron mobility.

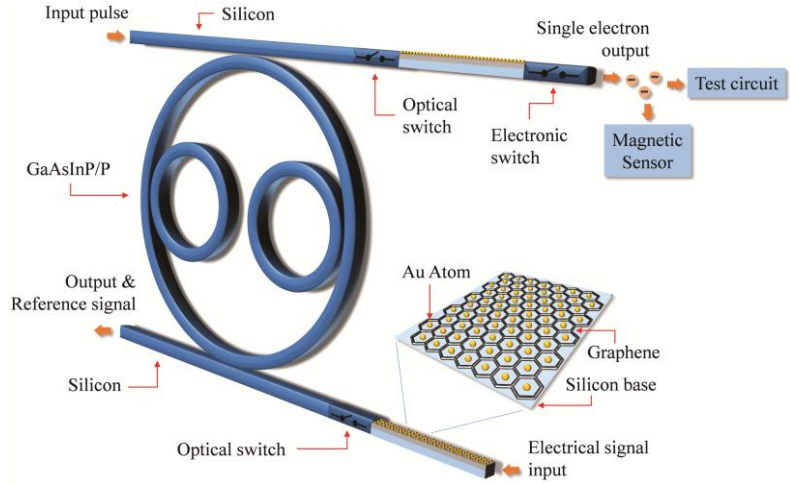

**Figure 2**: A schematic of electron mobility enhancement by driven electron speed control, where $R_1 = 400\ \mu m$, $R_2 = R_3 = 12\ \mu m$, the input pulse is a 1.55 $\mu m$, is a center wavelength of a Gaussian pulse, with a half width of $1.5 \times 10^{-14}$ sec.

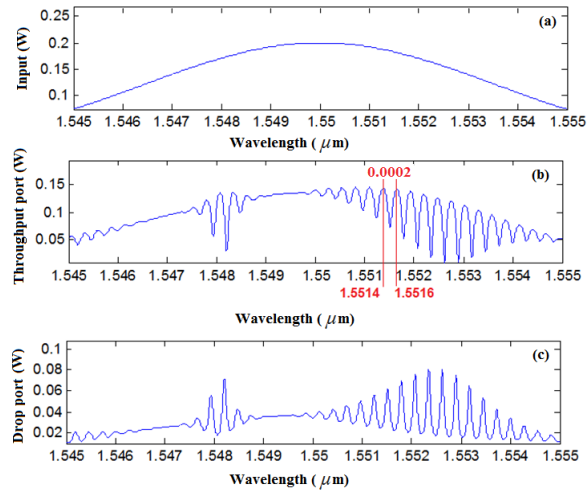

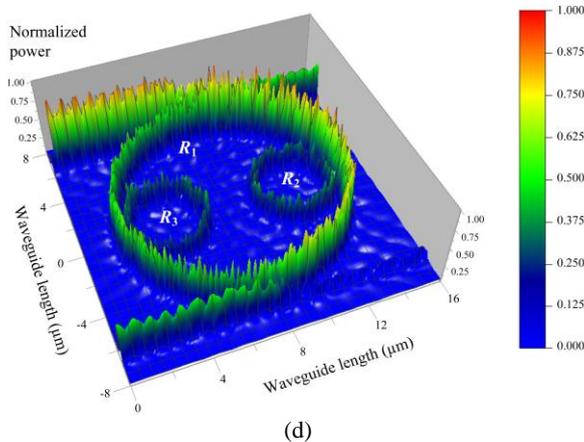

(d)

**Figure 3**: The enhanced electron mobility concept and system incorporating the stacked gold atom-dielectric-interfaces, the coupling constants are ranged between 0.50-0.52, where (a) the input, (b) the through and (c) the drop port electric field intensities, respectively, (d) the 3D imaging obtained by Opti-wave program, the GaAsInP/P refractive index is 3.14, the nonlinear refractive index is $2.7 \times 10^{-13}\ m^2W^{-1}$, the calculated group velocity obtained is around $v = \Delta v x \lambda_0\ cm\ s^{-1}$. From the reported [1], the electron mobility of silicon in graphene is $3 \times 10^7\ cm^{-2}$

$V^{-1}s^{-1}$ at low carrier density, in the calculation, the obtained group velocity is $\sim 2 \times 10^7\ cm\ m^{-1}$. Using $v = \frac{c}{n} \times \Delta\lambda$, where c is the spped of light in vaccum, n is the GaAs refractive index = 3.14, $\Delta\lambda$ = 0.2 nm, from Figure 3(b), with the frequency is femtosecond.

From Figure 2, the group velocity value can be evaluated by the space between two optical field peaks, where the higher group velocity is observed when the narrower peaks occur. The group velocity can be adjusted by changing the system parameters, where the major role is the two side ring radii, which can increase the speed of pulse repetition rate(switching in time), which can lead to obtaining the greater group velocity. The optical fields with group velocity enhancement is then entered into the stacked waveguide via the silicon base after the switching operation, from which the evanescent fields are penetrated into graphene atom, from which the electrons can be excited and conducted via the stacked gold atom, while the electron drift velocities are driven by the field group velocities and externally applied voltage(potential difference), which made the electron relative velocity much higher within the gold atom line than the drift velocity themselves. The relative electron mobility or permeability can be modified and significantly increased by using the gold atom or ferroelectric material as the conducted materials, where the use of such a devices and system for single electron and spintronic electronics can be realized.

In application, the upstream of the system of the operation is now described. On the other hands, the downstream conversion can be configured as following details. Principally, the Add port input can be used as the modulated port, where the external signals can be input and modulated into the system for wider applications. In addition, apart from modulation the Add port can also be used as the downstream input, from which the electron (current) can be input into the stacked device, where light can be generated by the electron density induced by the graphene atoms, eventually, the evanescent wave from graphene is transferred by the plasmonic waves and propagates within the silicon, then entering into the system, the output can be seen at the Drop port via the optical detector.

The use of the proposed concept can be interpreted into categories as followings. Firstly, the electron mobility of graphene on silicon without driven field if found in reference [5]. There are two important electron properties in electronics and magnetic materials, which are mobility and permeability, respectively. Firstly, the electron mobility (charge carrier) in a material is induced by electron drift velocity ($v$), the unit is $ms^{-1}$, the magnitude of the drift velocity is $v = \mu E$, $E$ is the electric field, $\mu = v/E$. Secondly, the electron permeability, is a constant of proportionality that exists between magnetic induction and magnetic field intensity, which is equal to $1.257 \times 10^{-6}$ $Hm^{-1}$ in free space, thus, $\mu = \sqrt{\frac{3}{2}} \times \frac{q}{m_e} \times \frac{h}{2\pi}$, where q and $m_e$ are the electron charge and mass, h is the Plank's constant. $\mu$ is the mangnetic moment. In a solid spin, of many electrons can act together to effect the magnetic and electric properties of a material. The use graphene on silicon base has been the promising form of a substrate that can be employed to serve the new era of electronics. Secondly, spintronic is the spin transport electronics, where the spin diffusion is the behaviors. The use of graphene on silicon is the very interesting aspect of researches and investigations. In Figure [16], to reflect the perfectly symmetric spin wave potential induced by the ring shape of the ferromagnet, we may completely eliminate the crystalline signature by preparing the amorphous ferromagnetic thin film. The amorphous ferromagnetic film is fabricated by ultra-high vacuum evaporation on an appropriate seed layer. The resonant property of the prepared film will be evaluated by the ferromagnetic resonance using a vector network analyzer. The damping constant of the prepared film will be less than 0.001, which is comparable to YIG film. By using ultra-high-precision electron-beam nanolithography, panda ring resonator consisting of perfectly circular ferromagnetic nano-rings will be fabricated. The width, thickness and its diameter designed by the numerical simulation will be realized. By the application of the local magnetic field and/or the spin injection, the spin wave in the ferromagnetic ring will be excited in the ferromagnetic ring. By measuring the magneto-resistance of the ferromagnetic ring under the microwave irradiation with various frequencies, where the standing spin-wave will be stabilized. In addition, the laser spot beam irradiation will be used for exciting the spin wave. The spin wave in the side rings of the panda ring resonator will be excited by the magneto-static interaction and/or spin-transfer effect from the center ring. Based on the theoretical prediction developed in 6.1, the nonlinear response of the excited spin wave excitation will be demonstrated experimentally.

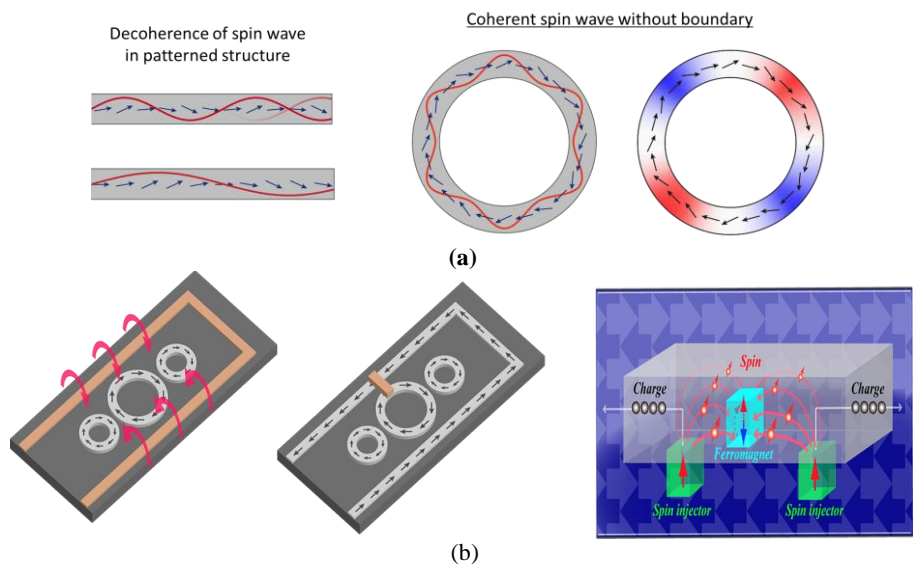

**Figure 4:** Schematic illustration of spin-based panda ring resonators, where (a) The schematic of (left) Edge boundary makes a de-coherence (evanescent) of spin wave, (middle) and (right) ferromagnetic nanoring without, (b) left one is field-driven operation and middle one is spin-current control, where the right hand graphic corresponds to the spin current

## 4. Conclusion

We have demonstrated that the electron mobility or permeability can be increased by using the driven optical field group velocity concept, where the electron speed can be relatively increased. By using the proposed system, the optical field group velocity can be increased by the ultra-switching pulse, which is obtained by external nonlinear coupling within the mirroring resonator system, in which the fast switching pulses in terms of group velocity is generated by the two nonlinear side rings, which is made by GaAsInP/P material. From the obtained results have shown that the increasing in electron speed in terms of driven group velocity, which is added up from the actually drifted velocity can be obtained, which can be useful for the hybrid electronic use, where the use of a single electron device for electronics and spintronics can be realized. For examples, a single electro device based on such a concept is the very interesting aspect of applications, while the ultra-sensitive magnetic field sensor using spin-based panda ring resonator, which is a technology based on the innovative principle conducted by the fusion of different research fields. It is no doubt to be a breakthrough technology on the extremely difficult research areas such as cyber-physical network and human healthcare devices. The realization of the high-performance wireless field sensor and remote chip, one of the ways for realizing the hardware with artificial intelligence, is a mankind's dream. Moreover, the achievement of spin-based panda ring resonator would break the performance limits of present semiconductor IC device and build out a leading technology in the world.

**Acknowledgment**

The authors would like to give the acknowledgement of the research facilities to Ton Duc Thang University, Ho Chi Minh City Vietnam.